\documentclass[aip,jap,reprint]{revtex4-1}

\usepackage{graphicx}
\usepackage{scrextend}

\begin{document}


\title{Temperature-dependent photo-response in multiferroic BiFeO\textsubscript{3} revealed by transmission measurements}



\author{F. Meggle}
\affiliation{Experimentalphysik 2, Universit\"at Augsburg, D-86159 Augsburg, Germany}

\author{M. Viret}
\affiliation{Service de Physique de l'Etat Condens\'e{}, DSM/IRAMIS/SPEC, CEA Saclay URA CNRS 2464, 91191 Gif-Sur-Yvette Cedex,
France.}

\author{J. Kreisel}
\affiliation{Physics and Materials Science Research Unit, University of Luxembourg, L-4422 Belvaux, Luxembourg.}
\affiliation{Department Materials Research and Technology, Luxembourg Institute of Science and Technology, 41 Rue du Brill, L-4422
Belvaux, Luxembourg.}

\author{C. A. Kuntscher}\email{christine.kuntscher@physik.uni-augsburg.de}
\affiliation{Experimentalphysik 2, Universit\"at Augsburg, D-86159 Augsburg, Germany}


\date{\today}

\begin{abstract}
We studied the light-induced effects in BiFeO$_3$ single crystals as a function of temperature by means of optical spectroscopy. Here we report the observation of several light-induced absorption features, which are discussed in terms of the photostriction effect and are interpreted in terms of excitons. The temperature dependence of their energy position suggests a possible coupling between the excitons and the lattice vibrations. Moreover, there are hints for anomalies in the temperature evolution of the excitonic features, which might be related to the temperature-induced magnetic phase transitions in BiFeO$_3$. Our findings suggest a coupling between light-induced excitons and the lattice and spin degrees of freedom, which might be relevant for the observed ultrafast photostriction effect in multiferroic BiFeO$_3$.
\end{abstract}



\maketitle

\section{Introduction}
In the last decades, bismuth ferrite BiFeO$_3$ (BFO) has attracted a significant interest due to a plethora of unexpected and exotic findings, related to outstanding magnetic, dielectric, and electric properties.\cite{Hill.2000,Yang.2009b,Yang.2010,Alexe.2011,Seidel.2011,Bhatnagar.2013} BFO belongs to the class of magnetoelectric mulitferroics, i.e., it shows simultaneously magnetic and electric orderings in the same phase, which are coupled to each other.\cite{Eerenstein.2006, Lebeugle.2008} Below $T_{\mathrm{N}}$=640~K the spins are aligned in an antiferromagnetic order and for temperatures below $T_{\mathrm{C}}$=1109~K BFO is ferroelectric.\cite{Ramazanoglu.2011, Kumar.2000} Due to the high transition temperatures BFO shows multiferroic behavior even at room temperature. Recently, it was shown that the dielectric and multiferroic properties of BFO improve significantly by doping BiFeO$_3$ with a small amount of nickel (i.e., BiFe$_{1-x}$Ni$_x$O$_3$ with $x\leq{}$0.07).\cite{Nadeem.2018}

An intensively studied but still little understood subarea of multiferroic materials is their interaction with light.\cite{Kreisel.2012, Paillard.2016a} As an example, BFO shows a photostriction effect, i.e., a coupling between light and the spatial dimensions of the illuminated crystal: Kundys et al.\cite{Kundys.2010} reported a small, but noticeable light-induced elongation of BiFeO$_3$ single crystals during illumination with white light from a normal bulb and red laser light ($\lambda{}$=632.8~nm). Photostriction in ferroelectrics is generally explained in terms of the combination of the bulk photovoltaic effect, where the incident light generates charges which screen the internal electric polarization and induce a change in the depolarization field, and the
inverse piezoelectric effect, which causes strain in the material.\cite{Lejman.2014,Kundys.2010, Paillard.2016a}

X-ray and optical pump-probe measurements on the photostriction effect in BFO found ultrafast response times in the range of nanoseconds or even faster.\cite{Li.2015, Schick.2014, Wen.2013, Chen.2012, Zhang.2014} In contrast, the photostriction response time for classic ferroelectric materials, such as lead lanthanum zirconate titanate (PLZT), amounts to tens of seconds.\cite{Kundys.2015} The ultrafast photostriction effect in BFO was attributed to the formation of excitons, i.e., charge-neutral electron-hole pairs, during light illumination and their diffusion and dissociation into screening charges.\cite{Li.2015} A light-induced absorption feature located at 540~nm ($\approx$2.3~eV) was associated with excited-state absorption of electrons in the conduction band and holes in the valence band, where the excited charge carriers might be trapped locally by defects, domain walls, or due to Debye screening.\cite{Wen.2013} Based on these findings, it was concluded that the photostriction effect in BFO is due to an electronic mechanism, rather than thermal effects.\cite{Wen.2013} Recent pump-probe measurements on the photovoltaic effect in epitaxial BiFeO$_3$ thin films
revealed a light-induced absorption feature at $\approx$2.3~eV and associated it with charge-neutral excitons, which diffuse and dissociate localized at sample interfaces.\cite{Li.2018} A general drawback of pump-probe measurements is their limitation to a rather narrow frequency range. An optical spectroscopy study\cite{Burkert.2016} covering a broad frequency range (1.1 -- 2.5~eV) found three light-induced absorption features in single-crystalline BFO during illumination with green laser light, which were associated with charge-transfer excitons or in-gap defect states.\cite{Pisarev.2009,Vikhnin.2002,Eglitis.2002}

Since the optical gap of BFO changes significantly with temperature\cite{Weber.2016}, temperature-induced changes in the photo-response provide very interesting extra information on the nature of the transitions.\cite{Zhang.2014} 
In this paper we investigate the influence of low temperature on the photo-response in single-crystalline BFO between 1.1~eV and 2.5~eV by optical spectroscopy. The results provide further information on the underlying mechanism of photostriction in multiferroic BiFeO$_3$.

\begin{figure}[t]
\includegraphics[width=0.75\columnwidth]{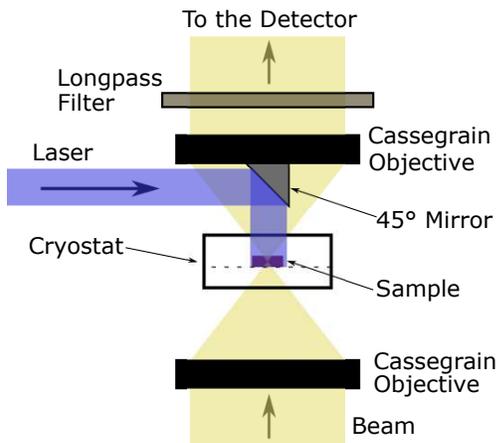}
\caption{Schematic presentation of the transmission experiment in the infrared microscope. The incoming beam is focused onto the sample inside the cryostat by the lower Cassegrain objective. The transmitted beam is collected by the upper Cassegrain objective and passes a longpass filter before it reaches the detector.}
\label{fig:setup}
\end{figure}

\section{Experimental Method}
The investigated BFO single crystal was grown by the flux method.\cite{Haumont.2008, Lebeugle.2007(b)} X-ray diffraction measurements revealed the high phase-purity of the sample, with only an additional Bi$_{25}$FeO$_{40}$ impurity phase of less than 0.9$\%{}$ in volume.\cite{Haumont.2008, Haumont.2009} The hexagonal unit cell parameters at ambient conditions were identified to $a_{\mathrm{hex}}=5.578(2)~\mathring{\mathrm{A}} $ and $c_{\mathrm{hex}}=13.865(3)~\mathring{\mathrm{A}}$ (see Ref. \citenum{Haumont.2009}), which is in good agreement with literature. \cite{Reyes.2007}
The sample has the lateral dimensions $0.35$~mm $\times{}$ 0.9~mm and was polished to a small platelet with thickness of $\approx$$39~\mu{}$m. Polarized transmission microscopy revealed its single-domain state (not shown).
For optical measurements in the temperature range from $295$~K to $20$~K the sample was mounted on a cold-finger Mikro-cryostat from
CryoVac Konti. The transmission measurements in the frequency range from $8500$~cm$^{-1}$ to $20000$~cm$^{-1}$ (1.05~eV to 2.48~eV) were performed with a Bruker Hyperion infrared micros\-cope, coupled to a Bruker Vertex80v FTIR spectrometer using a Si diode detector.
A blue MBL-III-473 laser from CNI ($\lambda{}=473$~nm, $E=2.6$~eV, $P=23.5$~mW, with polarization ratio larger than 100:1, cw) was used to illuminate the crystal.
To this end, the laser beam was deflected onto the sample via a $45^{\circ}$ aluminum mirror, which was glued on the upper Cassegrain objective of
the Hyperion microscope (see the scheme in Fig. \ref{fig:setup}).

In order to protect the Si diode from laser back scattering, a longpass filter ($\lambda_{\mathrm{cut-off}}=495$~nm) was inserted in
front of the detector.
The choice of the correct longpass filter is crucial, since its cut-off frequency sets the high-energy limit of the measurement range of the Si diode. Furthermore, the laser wavelength for illumination needs to be smaller than $\lambda_{\mathrm{cut-off}}$. Our choice of laser, longpass filter, and measurement range satisfies these requirements.

\begin{figure}[t]
\includegraphics[width=1\columnwidth]{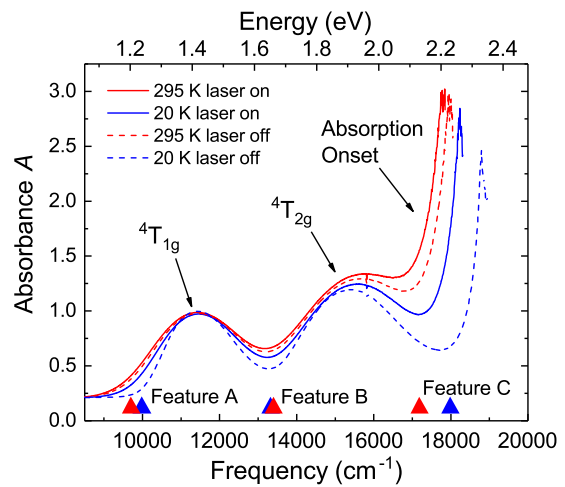}
\caption{Absorbance spectra of illuminated and non-illuminated BFO at $20$~K and $300$~K. The filled red and blue triangles mark the energy positions of the light-induced absorption features (labeled as Feature A, B and C) at 295~K and 20~K, respectively.}
\label{fig:absorbance_20and300kelvin}
\end{figure}

\section{Results and Discussion}

The influence of laser illumination and low temperature on the BFO absorbance spectrum is illustrated in Fig.\ \ref{fig:absorbance_20and300kelvin}. Here, the absorbance spectra of the BFO crystal are depicted at two selected temperatures (295~K and  20~K as the highest and lowest temperature of the cooling cycle, respectively) before and during illumination. The absorbance spectrum $A(\nu{})$ was calculated from the corresponding transmission spectrum $T_{\mathrm{off/on}}(\nu)$ according to $A(\nu{})=-\log_{10}{T_{\mathrm{off/on}}(\nu)}$. Hereby, the transmission spectrum $T_{\mathrm{off/on}}(\nu)$ was obtained according to the formula $T_{\mathrm{off/on}}(\nu)=I_{\mathrm{BFO,off/on}}(\nu{})/I_{\mathrm{ref}}(\nu{})$, where $I_{\mathrm{BFO,off/on}}(\nu{})$ is the intensity transmitted by the BFO crystal before (``off'') or during (``on'') laser illumination and  $I_{\mathrm{ref}}(\nu{})$ is
the intensity measured for the empty beam path.

All absorbance spectra show similar characteristics, namely two absorption bands related to the $d$-$d$ crystal field excitations ($^{6}\mathrm{A}_{1\mathrm{g}}\rightarrow{}^{4}\mathrm{T}_{1\mathrm{g}}$, $^{6}\mathrm{A}_{1\mathrm{g}}\rightarrow{}^{4}\mathrm{T}_{2\mathrm{g}}$) followed by a steep absorption onset due to excitations across the charge gap.
\cite{Ramachandran.2010,Neaton.2005,GomezSalces.2012,Xu.2009,Kumar.2008}
At room temperature and without illumination the crystal field excitations are located at around $1.41$~eV and $1.91$~eV, respectively (see Fig.\ \ref{fig:absorbance_20and300kelvin}) . This is in excellent agreement with previous optical measurements on BFO at ambient conditions.\cite{Burkert.2016, Xu.2009, GomezSalces.2012}
During illumination the room-temperature absorbance spectrum is enhanced due to additional excitations in three specific energy ranges, namely at around $1.2$~eV, $1.7$~eV, and $2.1$~eV, consistent with previous observations.\cite{Burkert.2016} The light-induced absorption changes become even more pronounced at $20$~K as depicted in Fig.\ \ref{fig:absorbance_20and300kelvin}.

\begin{figure}
\includegraphics[width=1\columnwidth]{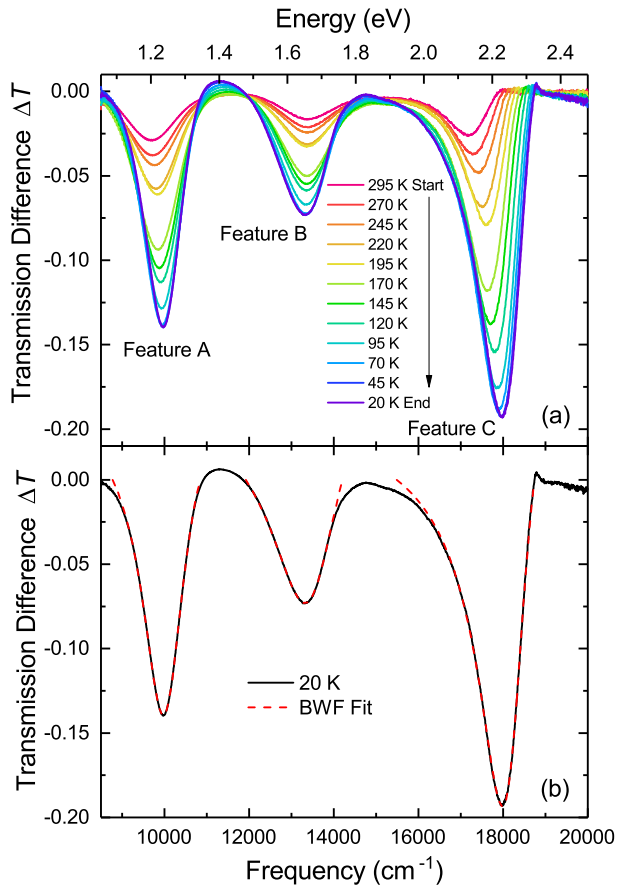}
\caption{(a) Transmission difference $\Delta{}T(\nu{})$ showing the absorption features A, B, and C in the temperature range between 295~K and 20~K.
(b) Transmission difference at $20$~K together with the BWF fit for the three absorption features. The gray bars in (a)-(b) indicate the temperature ranges for hints of anomalies.}
\label{fig:photostriction_spectrum_all}
\end{figure}

For a quantitative identification of these photo-induced spectral changes we consider the transmission difference
$\Delta{}T(\nu)$ calculated as
\begin{equation}
\Delta T(\nu) = (I_{\mathrm{BFO,on}}(\nu) - I_{\mathrm{BFO,off}}(\nu))/I_{\mathrm{ref}}(\nu) \quad .
\label{eq:Delta_trans_norm}
\end{equation}
The so-obtained transmission difference in steps of $25$~K during cooling down from 295 to 20~K is depicted in Fig.\ \ref{fig:photostriction_spectrum_all}(a).
Each spectrum consists of three asymmetric absorption features, which are labeled as feature A, B, and C in the following.
With decreasing temperature the absolute intensities increase for all three features and the features become sharper, likely due to the expected reduced thermal broadening. Furthermore, during cooling down the energy positions of the features behave in a distinct way: while features A and C shift to higher energies, i.e., show a blue shift, with decreasing temperature, feature B undergoes a red shift.

\begin{figure}[t]
\includegraphics[width=1\columnwidth]{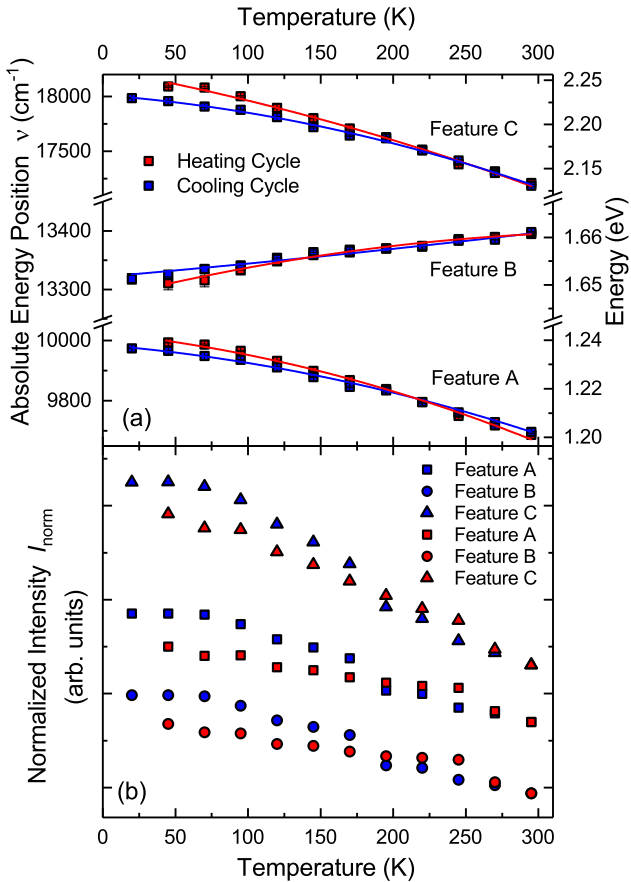}
\caption{(a) Absolute energy positions of the three absorption features A, B, and C as a function of temperature during the cooling cycle (blue points) and during the heating cycle (red points), with second-order polynomial fits as guides to the eye. (b)  Normalized intensity $I_{\mathrm{norm}}$ of the three absorption features A, B, and C as a function of temperature
(blue symbols: cooling cycle, red symbols: heating cycle). For
clarity the normalized intensities are plotted with an offset.}
\label{fig:position_absolute}
\end{figure}

In order to determine their energy positions and intensities, the absorption features were fitted with an asymmetric Breit-Wigner-Fano (BWF) lineshape with the formula
\begin{equation}
I_{\mathrm{BWF}}(\nu{}) = I_0 +
H\left(1+\frac{\nu{}-\nu{}_{\mathrm{c}}}{q\Gamma{}}\right)^2/\left(1+\left(\frac{\nu{}-\nu{}_{\mathrm{c}}}{\Gamma{}}\right)^2\right).
\label{BWF}
\end{equation}
Here, $\nu{}_{\mathrm{c}}$, $1/q$, $\Gamma{}$, $H$, and $I_0$ correspond to the central frequency, asymmetry factor, spectral width, intensity factor, and offset, respectively.
As an example, we show in Fig.\ \ref{fig:photostriction_spectrum_all}(b) the BWF fit of the three absorption features at $20$~K.
From the fit parameters one can calculate the actual energy position of the absorption features.
To this end, the first derivative of the BWF function is set to zero, yielding the energy position
$\nu_{i} = \Gamma{}/q+\nu{}_{\mathrm{c}}$ for $i$=A,B,C. Accordingly, from the BWF fitting we obtained
the feature positions $\nu{}_{\mathrm{A}}=9697$~cm$^{-1}$ ($E_{\mathrm{A}}=1.20$~eV),
$\nu{}_{\mathrm{B}}=13398$~cm$^{-1}$ ($E_{\mathrm{B}}=1.66$~eV) and $\nu{}_{\mathrm{C}}=17186$~cm$^{-1}$ ($E_{\mathrm{C}}=2.13$~eV)
at room temperature.
This is in good agreement with earlier room-temperature measurements.\cite{Burkert.2016} The energies of the absorption features clearly differ from the energies of the $^{4}\mathrm{T}_{1}$ and $^{4}\mathrm{T}_{2}$ crystal field excitations and the absorption onset in BFO (see Fig. \ref{fig:absorbance_20and300kelvin}). For illustration, the positions of the absorption features under laser illumination relative to the crystal field excitations and the absorption onset are indicated by filled triangles in Fig.\ \ref{fig:absorbance_20and300kelvin}.
The asymmetry of the absorption features, which is most obvious for feature C, arises from their close vicinity to the intrinsic excitations in BFO (absorption onset and crystal field excitations). The temperature-dependent energy positions of features A, B, and C as obtained from the fittings are shown in Fig.\ \ref{fig:position_absolute}(a).

\begin{figure}
\includegraphics[width=0.95\columnwidth]{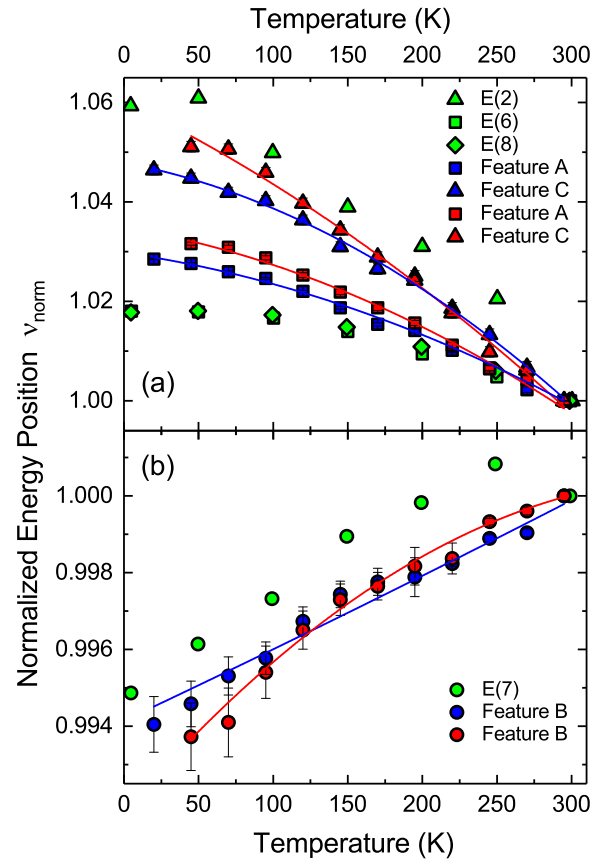}
\caption{Normalized energy positions $\nu{}_{\mathrm{norm}}$ of the three absorption features A, B, and C as a function of temperature (blue symbols: cooling cycle, red symbols: heating cycle) are depicted in (a) and (b) together with the normalized frequencies of selected infrared-active phonon modes (green symbols) from Ref.\ \citenum{Lobo.2007}.}
\label{fig:position_intensity_normalized}
\end{figure}

In order to compare the temperature behavior of the three features, we normalized the energy position by its room-temperature value according to $\nu{}_{i\mathrm{,norm}}(T)$=$\nu{}_{i}(T)/\nu{}_{i}(T=295~\mathrm{K})$
for $i$=A, B, and C. The so-obtained normalized energy positions of the three features as a function of temperature are depicted in Figs.\ \ref{fig:position_intensity_normalized}(a) and \ref{fig:position_intensity_normalized}(b).
As already mentioned above, the absorption features A and C harden, whereas feature B softens during cooling. In addition, features A and C shift much stronger with temperature than feature B. In the cooling cycle (blue symbols in Figs. \ref{fig:position_absolute}(a), \ref{fig:position_intensity_normalized}(a), and \ref{fig:position_intensity_normalized}(b)) there are weak indications for an anomaly between 195~K and 170~K. During warming up (red symbols) one notices a weak anomaly at around 245~K and a saturation for temperatures below 70~K.

The intensity of each feature was calculated according to Equ.\ (\ref{BWF}) by inserting all the extracted parameters at $\nu_{i}$ and then normalized to its value at 295~K. The so-obtained normalized intensity $I_{\mathrm{norm}}$ as a function of temperature is plotted in Fig.\ \ref{fig:position_absolute}(b). The intensity of all three features increases during cooling, and in addition one can see anomalies in intensity as well. The intensities in the cooling cycle (blue symbols) show a sudden increase between 195~K and 170~K and saturate below 70~K. During warming up (red symbols) there is an abrupt decrease in intensity at 245~K observable and there are hints for an intensity saturation between 20~K and 95~K.

The temperature dependence of the absorption features induced by laser illumination provides insight into the mechanism underlying the photostriction effect in BFO. The photostriction effect is explained in terms of a combination of photovoltaic and inverse piezoelectric effect, most probably involving the formation of excitons. However, the detailed microscopic mechanism underlying photostriction in BFO is still under controversial discussion.
Early reports supposed that the incident light creates electron-hole pairs which are separated by the ferroelectric polarization in BFO.\cite{Choi.2009, Bhatnagar.2013} The charge carriers migrate to the surface in order to screen the internal depolarizing fields. This photovoltaic voltage creates a strain due to the inverse piezoelectric effect, which leads to photostriction. 
The results of recent time-resolved x-ray measurements on BFO thin films suggested the formation of electron-hole pairs, i.e., charge-neutral excitons, during illumination.\cite{Li.2015} The surface-located excitons dissociate due to local band bending and the free carriers screen the depolarization field; due to the inverse piezoelectric effect an immediate strain is generated. Furthermore, an earlier room-temperature optical study on the photostriction effect in BFO found three absorption features during illumination, which were interpreted in terms of charge-transfer vibronic excitons or in-gap defect states.\cite{Burkert.2016} Such excitons involve charge excitations between $p$ and $d$ states and could be influenced by the vibronic interactions with the underlying lattice.\cite{Pisarev.2009,Eglitis.2002,Vikhnin.2002}

Temperature-dependent resonant Raman scattering experiments found evidence for in-gap states due to oxygen vacancies with excitations energy at around 2.4~eV.\cite{Weber.2016} Also cathodoluminescence measurements on BFO thin films found additional absorption features at 2.20 and 2.45~eV and interpreted them as defect states.\cite{Hauser.2008} The absorption feature C in our data occurs at $E_{\mathrm{C}}=2.13$~eV, i.\ e., it is close in energy. However, its rather strong temperature dependence is inconsistent with an interpretation in terms of defect states.

\begin{figure}[t]
\includegraphics[width=0.8\columnwidth]{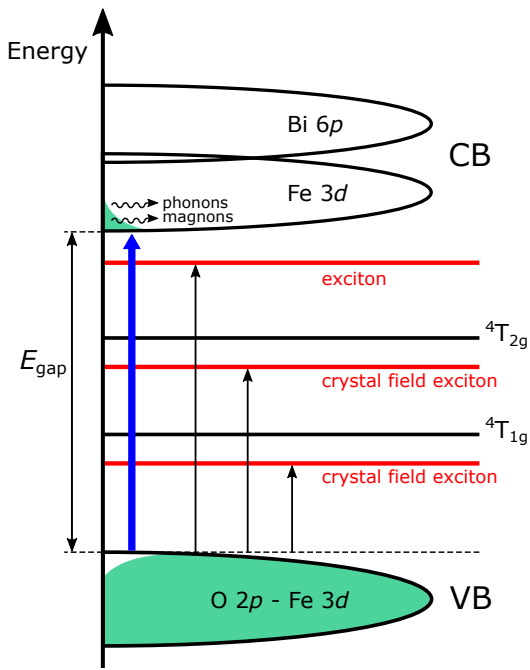}
\caption{Scheme of the electronic band structure of BFO, in analogy to Ref.\ \citenum{Chen.2012}, with additional excitonic states (marked in red) appearing during illumination. The valence band (VB) has O $2p$ - Fe $3d$ character and the conduction band (CB) has Fe 3$d$ and Bi 6$p$ character.\cite{Wang.2009(b), Higuchi.2008} Occupied electronic states in the VB and CB during laser illumination are marked by the green area. The blue arrow indicates the charge excitations induced by the incident laser radiation.
The three black arrows illustrate the transitions associated with the absorption features A, B, and C as depicted in Fig.\ \ref{fig:photostriction_spectrum_all}.}
\label{fig:exciton-drawing}
\end{figure}

Therefore, an interpretation of the observed absorption features during illumination in terms of excitons \cite{Li.2015,Pisarev.2009}, with a possible coupling of the involved electronic levels with other degrees of freedom - such as phonons \cite{Vikhnin.2002, Eglitis.2002} and magnons \cite{Xu.2009} - shall be considered. Indeed, the generation of phonons in BFO during light illumination was demonstrated recently.\cite{Chen.2012,Lejman.2014}
Also, the importance of spin-charge-lattice coupling in multiferroics was emphasized.\cite{Ramirez.2009}
Since the three features A, B, and C appear at energies 0.15 - 0.25~eV below the crystal field excitations and the absorption onset, respectively, we associate the involved excitons to the corresponding electronic levels in BFO. The suggested electronic band scheme of BFO under illumination is illustrated in Fig.\ \ref{fig:exciton-drawing}. It consists of the valence band (VB) due to O $2p$ - Fe $3d$ states, the conduction band (CB) due to Fe 3$d$ and Bi 6$p$ states, and the two crystal field excitations $^{4}\mathrm{T}_{1\mathrm{g}}$ and $^{4}\mathrm{T}_{2\mathrm{g}}$. Electrons excited by the incident laser radiation into the CB release energy through the emission of phonons or magnons.

During laser illumination three additional states (marked in red in Fig.\ \ref{fig:exciton-drawing}, presumably excitons, appear between the valence and the conduction band.
The two lower-lying exciton levels are associated with the crystal field excitations. A so-called crystal field exciton generally describes a crystal field excitation propagating from site to site due to exchange interaction.\cite{Raymond.2014} We use the expression here in the broader sense of an exciton related to the crystal field excitation coupled with the spin or lattice degrees of freedom.
The highest-lying exciton level could be associated with either charge-transfer excitations or higher-energy crystal field excitations.\cite{Ramirez.2009} We thus associate the three absorption features A, B, and C with excitations into the three excitonic levels, as illustrated by the three black arrows in Fig.\ \ref{fig:exciton-drawing}.

In order to test a possible coupling of the light-induced excitonic features with the lattice degrees of freedom, we compared the temperature dependence of their energy position with that of the phonon modes in BFO. The temperature dependence of the infrared-active phonon modes has been studied in detail by several groups in the last years. \cite{Lobo.2007, Rovillain.2009, Palai.2010, Haumont.2006b} Group theory predicts for $R3c$ symmetry 18 phonon modes in total, which can be decomposited in 4$A_1$+5$A_2$+9$E$ modes.\cite{Hermet.2007} The $A_1$ and the $E$ modes are polarized along the $c$ axis, and in the $x-y$ plane, respectively. Both are infrared and Raman active, whereas $A_2$ phonons are silent. Polarized infrared spectroscopy by Lobo et al.\ \cite{Lobo.2007} revealed nine $E$ and four $A_1$ phonon modes in good agreement with group theory. The frequencies of the $E$ phonon modes are summarized in Tab.\ I and Fig.~2 of Ref.\ \citenum{Lobo.2007}. At room temperature the frequency of the $E$ modes ranges from 66~cm$^{-1}$ to 521~cm$^{-1}$. From theoretical calculations it is known that the two low-energy modes ($E$(1) and $E$(2)) are ascribed to Bi atom motions, the six high energy modes ($E$(4) to $E$(9)) are mainly due to oxygen motions, and the remaining phonon mode ($E$(3)) is ascribed to iron motions.\cite{Hermet.2007} The nine $E$ phonon modes shift individually during a heating cycle from 5 to 300~K in their frequency positions: The phonon modes $E$(1), $E$(2), $E$(6), and $E$(8) move to lower frequencies with increasing temperature, $E$(3), $E$(4), $E$(5), and $E$(9) show almost no temperature dependence, and $E$(7) shifts to higher frequencies.

For comparing the temperature dependence of the light-induced absorption features with that of the phonon modes observed by Lobo et al.\ \cite{Lobo.2007}, we extracted the temperature-dependent phonon positions and normalized them to their energy position at room temperature. Figures\ \ref{fig:position_intensity_normalized}(a) and \ref{fig:position_intensity_normalized}(b) show the normalized frequency positions of the most promising phonon modes (green symbols) with those of the light-induced absorption features as a function of temperature. The similarity in temperature dependence is striking and strongly suggests that the temperature behavior of features A, B and C translated the coupling to phonons. The absorption feature A is close to the phonon mode $E$(2) over the whole temperature range, feature B shifts similarly to $E$(7), and feature C behaves close to modes $E$(6) and $E$(8).\footnote{The relatively strong temperature dependence of feature C could also stem from the rather strong temperature dependence of the absorption onset, in contrast to the crystal field excitations which are only weakly affected by temperature changes (see Fig. \ref{fig:absorbance_20and300kelvin})} This finding is in agreement with a possible coupling between the absorption features and the lattice vibrations in BFO.

It is worth to mention that earlier photoelastic stress measurements on BFO showed that the photostriction effect depends on magnetic fields.\cite{Kundys.2010} Thus, it is highly likely that the light-induced absorption features also couple to magnetism. Such a coupling is also expected from the magnetoelectric property of BFO. The temperature hysteresis of the feature positions below 200~K might be due to spin glass behavior of BiFeO$_3$ which was reported to start between 200~K and 160~K.\cite{Nakamura.1993, Pradhan.2005} If the hints for anomalies in our data at around 245~K, 195~K, and 70~K are real, they could be due to a coupling to spin reorientations, as well. In the literature one can find further magnetic transitions in BFO while cooling from room temperature, however, the transition temperatures are not always consistent.\cite{Scott.2008, Redfern.2008, Cazayous.2008, Zhang.2014, Singh.2008, Rovillain.2009} At around 240~K, 200~K, and 90~K spin-reorientation and anomalies in the temperature evolution of the energy and intensity of several magnon modes were found for BFO, which would match the temperatures of the anomalies observed for the positions and intensities of features A--C.

Due to the facts that (i) the temperatures of the anomalies observed for the light-induced features might be associated with the temperatures of the magnetic transitions and that (ii) the photostriction effect depends on the magnetic field, our data support a coupling between the light-induced absorption features and the spin degrees of freedom in BFO. Since the reports in the literature are incomplete regarding the temperature dependence of all magnon modes in BFO, we cannot specify to which magnons the light-induced features possibly couple.\cite{Cazayous.2008,Rovillain.2009,Singh.2008} It is interesting to note that a coupling between the on-site crystal-field excitations and magnetic excitations was already proposed in Ref. \citenum{Xu.2009}. In contrast to vibronic excitations, spin excitations are sensitive to an applied magnetic field. Thus, a suitable test for the coupled exciton-magnon excitations in BFO, as proposed here, are corresponding transmission measurements under laser illumination in an external magnetic field.

\section{Conclusion}
In conclusion, we studied the temperature-dependent transmission spectrum in the visible frequency range on single-crystalline BFO under continuous illumination. Three light-induced absorption features appear below the two crystal field excitations and the absorption onset, respectively. We suggest that the features correspond to (crystal field) excitons, which appear in the electronic band scheme of BFO during illumination. The temperature dependence of the energy positions and intensities of the light-induced absorption features suggest a coupling of the (crystal field) excitons to phonons and potentially also to magnons in BFO. Furthermore, an interpretation of the absorption features in terms of in-gap defect states seems less likely due to their rather strong temperature dependence.

\begin{acknowledgments}
We thank F.\ Burkert for technical support and fruitful discussions.
This work is financially supported by the Deutsche Forschungsgemeinschaft (DFG) through grant no.\ KU 1432/9-1. JK acknowledges financial support from the Fond National de Recherche Luxembourg through a PEARL grant (FNR/P12/4853155/Kreisel).
\end{acknowledgments}

\end{document}